\documentclass[aps,prl,twocolumn,showpacs,superscriptaddress]{revtex4-1}  
\pdfoutput=1
\usepackage{times}  
\usepackage[pdftex]{graphicx}
\usepackage{graphics}
\usepackage{dcolumn}   
\usepackage{bm}        
\usepackage{amssymb}   
\usepackage{siunitx}   
\usepackage{amsmath}
\usepackage{amssymb}
\usepackage{float}
\begin{document}

\title{Toward Scalable Boson Sampling with Photon Loss}
\author{Hui Wang}
\author{Wei Li}
\author{Xiao Jiang}
\author{Y.-M. He}
\author{Y.-H. Li}
\author{X. Ding}
\author{M.-C. Chen}
\author{J. Qin}
\author{C.-Z. Peng}
\affiliation{%
Shanghai Branch, National Laboratory for Physical Sciences at Microscale,
    	University of Science and Technology of China, Shanghai 201315, China%
	}
\affiliation{%
CAS Center for Excellence and Synergetic Innovation Center in Quantum Information and Quantum Physics, University of Science and Technology of China, Hefei, Anhui 230026, China%
    }
\affiliation{%
CAS-Alibaba Quantum Computing Laboratory, Shanghai 201315, China%
    }
\author{C. Schneider}
\author{M. Kamp}
\affiliation{%
Technische Physik, Physikalisches Instit{\"a}t and Wilhelm Conrad R{\"o}ntgen-Center for Complex Material Systems, Universitat W{\"u}rzburg, Am Hubland, D-97074 W{\"u}zburg, Germany%
    }
\author{W.-J. Zhang}
\author{H. Li}
\author{L.-X. You}
\author{Z. Wang}
\affiliation{%
State Key Laboratory of Functional Materials for Informatics, Shanghai Institute of Microsystem and Information Technology (SIMIT), Chinese Academy of Sciences, 865 Changning Road, Shanghai 200050, China%
    }

\author{J. P. Dowling}
\affiliation{%
Hearne Institute for Theoretical Physics and Department of Physics and Astronomy, Louisiana State University, Baton Rouge, Louisiana 70803, USA%
    }
\affiliation{%
NYU-ECNU Institute of Physics at NYU Shanghai, Shanghai 200062, China%
    }

\author{S. H{\"o}fling}
\affiliation{%
Shanghai Branch, National Laboratory for Physical Sciences at Microscale,
    	University of Science and Technology of China, Shanghai 201315, China%
    }
\affiliation{%
Technische Physik, Physikalisches Instit{\"a}t and Wilhelm Conrad R{\"o}ntgen-Center for Complex Material Systems, Universitat W{\"u}rzburg, Am Hubland, D-97074 W{\"u}zburg, Germany%
    }
\affiliation{%
SUPA, School of Physics and Astronomy, University of St Andrews, St Andrews KY16 9SS, United Kingdom%
    }
\author{Chao-Yang Lu}
\author{Jian-Wei Pan}
\affiliation{%
Shanghai Branch, National Laboratory for Physical Sciences at Microscale,
    	University of Science and Technology of China, Shanghai 201315, China%
	}
\affiliation{%
CAS Center for Excellence and Synergetic Innovation Center in Quantum Information and Quantum Physics, University of Science and Technology of China, Hefei, Anhui 230026, China%
    }
\affiliation{%
CAS-Alibaba Quantum Computing Laboratory, Shanghai 201315, China%
    }
\date{\today}

\begin{abstract}
Boson sampling is a well-defined task that is strongly believed to be intractable for classical computers, but can be efficiently solved by a specific quantum simulator. However, an outstanding problem for large-scale experimental boson sampling is the scalability. Here we report an experiment on boson sampling with photon loss, and demonstrate that boson sampling with a few photons lost can increase the sampling rate. Our experiment uses a quantum-dot-micropillar single-photon source demultiplexed into up to seven input ports of a 16$\times$16 mode ultra-low-loss photonic circuit, and we detect three-, four- and five-fold coincidence counts. We implement and validate lossy boson sampling with one and two photons lost, and obtain sampling rates of \SI{187}{kHz}, \SI{13.6}{kHz}, and \SI{0.78}{kHz} for five-, six- and seven-photon boson sampling with two photons lost, which is 9.4, 13.9, and 18.0 times faster than the standard boson sampling, respectively. Our experiment shows an approach to significantly enhance the sampling rate of multiphoton boson sampling.
\end{abstract}

\maketitle

Boson sampling \cite{AA2010BS} is considered as a strong candidate to demonstrate quantum computational supremacy \cite{Preskill2012, Harrow2017}. It only requires indistinguishable single-photon sources, a passive linear network, and single-photon detection. However, it is strongly believed to be intractable for classical computers under some computational complexity assumptions. Its relatively simple design attracts a number of proof-of-principle experiments \cite{Broome2013science,Spring2013science,Tillmann2013NP,Crespi2013NP},  using probabilistic heralded single photons produced by spontaneous parametric down-conversion (SPDC) \cite{Kwiat1995PRL}. Recently, high-performance single-photon sources based on quantum-dot micropillars were developed \cite{Ding2016PRL,Wang2016PRL,Smoaschi2016NP,Unsleber2016OE} and applied to multiphoton boson sampling, which significantly increased the photon number and sampling rate \cite{Wang2017NP,He2017PRL,Loredo2017PRL}.

\begin{figure*}[htbp]
  \centering
  \includegraphics[width=1\textwidth]{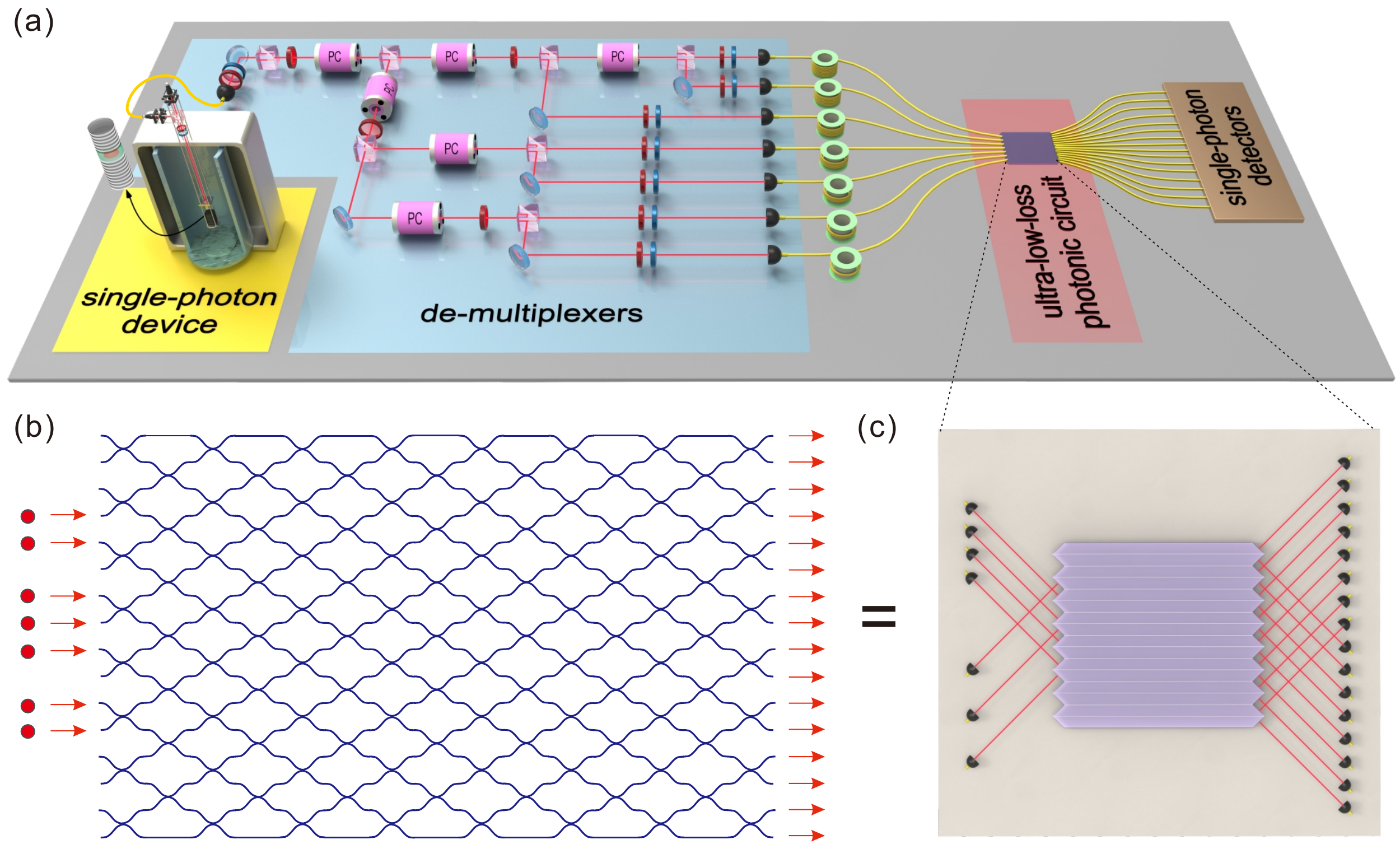}
  \caption{(a) Experimental setup for lossy boson sampling. The setup contains four parts. The first part is a single-photon source from a quantum-dot micropillar. It is placed inside a \SI{4.2}{K} cryostat, and a confocal microscopy is used to excite the quantum dot and collect its resonance fluorescence. The second part is six cascaded demultiplexers that separate the single photon stream into seven different spatial modes. Seven single-mode fibers with different lengths are used to compensate the time delay among seven different modes. The third part is the ultra-low-loss photonic network; the demultiplexed single photons are injected into a $16\times16$ mode square-shaped photonic network, which contains 113 beam splitters and 14 mirrors. The last part is the detection; 13 superconducting nanowire single-photon detectors and 3 silicon-based avalanche detectors are used to detect photons, and a home-made coincidence-count unit registers all no-collision events (not shown). (b) The equivalent photonic circuit of our $16\times16$ mode interferometer, which is fully-connected and has a transmission rate above 99$\%$. (c) Enlarged ultra-low-loss photonic network with a size of 50.91$\,$mm$\times$45.25$\,$mm$\times$4.00$\,$mm.}
  \label{fig:1}

\end{figure*}

In the photonic experiments, the major obstacle to scaling up, is the unavoidable photon loss, which can happen in the source, interferometer, and detectors. Recently, Aaronson and Brod have investigated boson sampling with photons lost \cite{Aaronson2016PRA}. In standard boson sampling, we send $n$ single photons into an $m$-port $(n$$\ll$$m)$ passive linear network, and then sample from the output distribution. Although serious losses in the single-photon source, linear networks, and detectors, we can post-select $n$-fold coincidence counts and compare to the theoretical distribution given by calculating permanents of sub-matrices. However, in the new variation of the scheme of boson sampling---lossy boson sampling, the only difference is that we post-select $(n-k)$-fold coincidence counts, where the $k$ is the number of lost photons.

Aaronson and Brod have shown that, if $k$ is constant, lossy boson sampling cannot be simulated in classical polynomial time, just under exactly the same complexity assumptions used for standard boson sampling \cite{Aaronson2016PRA}. This theoretical work indicates that boson sampling is a very robust model under experimental imperfections. Importantly, in the lossy scenario, the sampling rate can exponentially grow with $k$, which can make boson sampling more feasible in order to demonstrate quantum supremacy. Here, we experimentally investigate the first lossy boson sampling using a quantum-dot single-photon source with a system efficiency of 33.7$\%$ \cite{Wang2017NP}, a 16$\times$16 mode ultra-low-loss ($<$1$\%$) photonic network \cite{Reck1994PRL,Clements2016}, and inefficient single-photon detectors with an average efficiency of 53$\%$. Ref.  \cite{Aaronson2016PRA} only considered random path-independent loss that happens at the single-photon source. Here we give a result on a more realistic model that losses happen anywhere except the interferometer \footnote{See Supplemental Material for more information on the single-photon sources, demultiplexers, fabrication and characterize of the photonic networks, analysis of the lossy boson sampling, validation and experimental results of boson sampling with two photons lost.}.

\begin{figure*}
  \centering
  \includegraphics[width=0.88\textwidth]{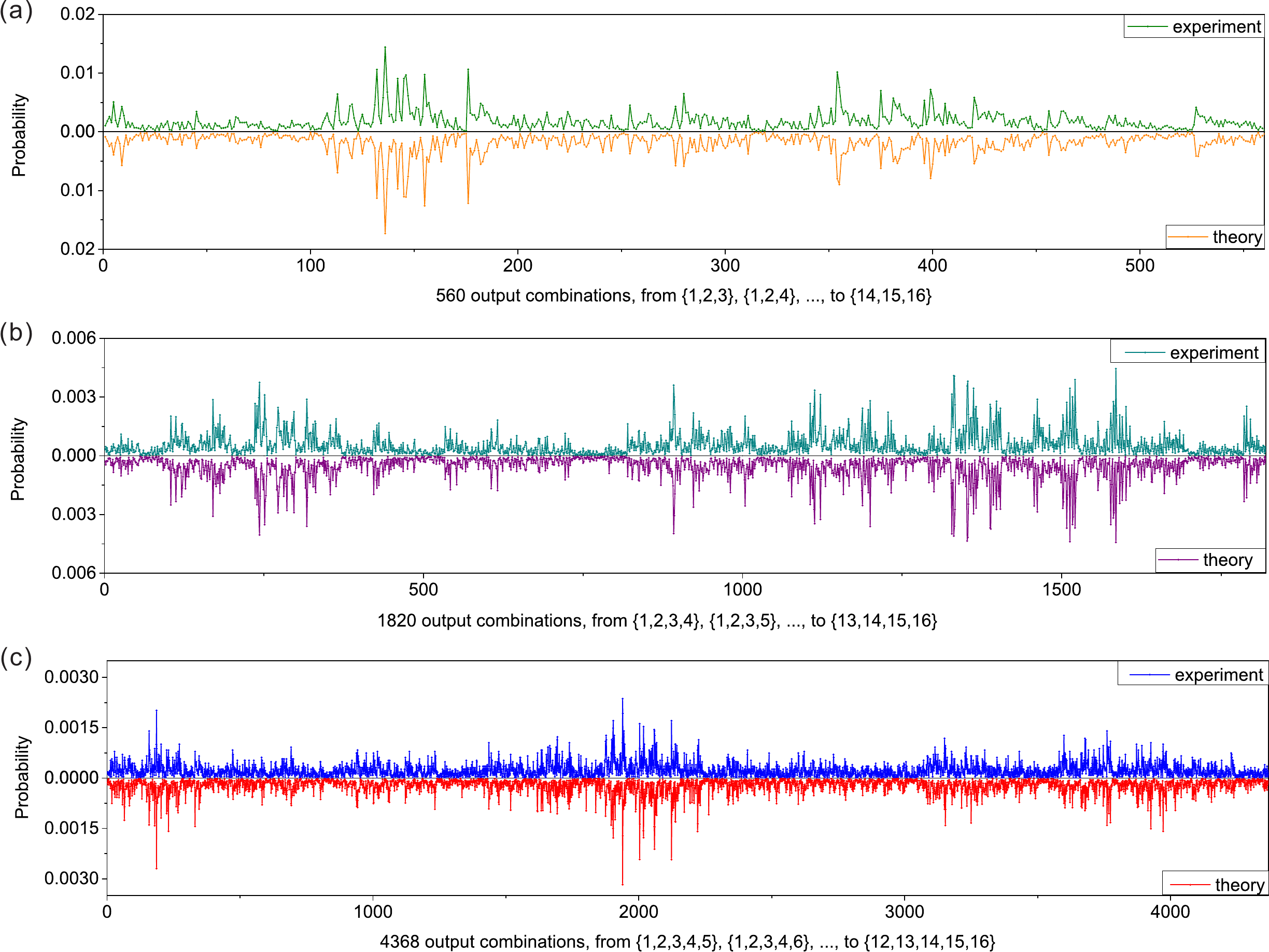}
  \caption{Experimental results for four-, five- and six-photon boson sampling with one photon lost. The upper part of each graph are the experimental results, and the bottom part are the theoretical results, given by calculating permanents. All no-collision output combinations are denoted by $\{i, j, \cdots\}$ where $i$, $j$,  is the $i$-th, $j$-th, $\cdots$ output port.  There are $560$, $1820$ and $4368$ output combinations for the four- (a), five- (b), and six-photon (c) boson sampling with one photon lost; the measured distances are 0.085(1), 0.106(2) and 0.201(3), and measured similarities are 0.994(1), 0.989(2) and 0.960(4), respectively.}
  \label{Fig:2}
\end{figure*}

As shown in Fig.$\,$\ref{fig:1}, a single self-assembled InAs/GaAs QD was embedded inside a micropillar cavity with a diameter of \SI{2}{\micro\metre}. Under $\pi$ pulse excitation \cite{YMHe2013nn} with a repetition rate of \SI{76}{MHz}, the QD-micropillar emits \SI{25.6}{MHz} polarized single photons at the end of the single-mode fiber, which are directly used for boson sampling without any spectral filtering \cite{Wang2017NP}. The measured second-order correlation function $g^2(0)$ and photon indistinguishability are 0.027(1) and 0.939(3), respectively. The high quality of single photons benefits from a large Purcell enhancement (a factor of $7.6$) of the microcavity which enhances the radiative rate \cite{Wang2016PRL}. The cavity Q factor is $\sim$6400 which is helpful to suppress the phonon sidebands~\cite{Smith2017NP}.

We actively demultiplexed the single-photon pulse stream into seven different spatial modes (see Fig.~\ref{fig:1} and \cite{Note1}). The demultiplexers consist of six cascaded Pockels cells and polarizing beam splitters (PBSs). Driven by half-wave voltage ($\sim$\SI{1800}{V}), a Pockels cell (with its axis aligned at $\pi/4$) can rotate the polarization by 90$^\circ$, and a PBS is used to separate the single-photon stream. With the demultiplexers operated at a repetition rate of \SI{0.775}{MHz}, every 98 pulses are separated into seven segments equally. That is, 14 sequential pulses will go out from each of seven different spatial modes \cite{Note1}. Then we use seven single-mode fibers with different lengths to ensure that each segment of the pulse trains arrive at interferometer simultaneously. Translation stages were used to finely adjust the arrival time with a precision of \SI{0.03}{ps}. Owing to the high transmission rate ($>$99$\%$) of the Pockels cells, and the high single-mode fiber coupling efficiency (92$\%$), the average efficiency of the demultiplexers is $\sim$85$\%$.

The single photons are fed into a 16$\times$16 mode square-shaped photonic network \cite{Clements2016}, which has the feature of high stability, ultra-low loss, and matrix randomness (see Fig.$\,$\ref{fig:1} and \cite{Note1}). Comparing to the triangle-shaped design by Reck et al. \cite{Reck1994PRL}, this square-shaped design achieves minimal optical depth, which requires less beamsplitters and phase shifters, and has less optical losses. On the other hand, this symmetry design is more robust to the optical losses \cite{Clements2016}. It has a size of 50.91$\,$mm$\times$45.25$\,$mm$\times$4.00$\,$mm which was fabricated by bonding 16 tiny trapezoids together. Every surface between a pair of trapezoids are optically coated with polarization-dependent beam-splitting thin films, while the top and bottom surfaces were total-reflecting coated. This network equivalently contains 113 beam splitters and 14 mirrors, which acts as a unitary transformation to the input Fock states. Note that our photonic circuit has a negligible loss (transmission rate $>$99$\%$). In this case, it is reasonable to consider it as a unitary matrix, which avoids the complex computation caused by path-dependent loss in the photonic network.

Thirteen superconducting nanowire single-photon detectors and three silicon-based avalanche detectors are used to detect the photons, and a 64-channel home-made coincidence counting unit is applied to register all no-collision events. We classify losses in the photonic paths into two groups: loss at the source---all losses before the photonic network, and loss at the detectors---all losses after the photonic network. Imagine that we have $n+k$ input ports in front of photonic circuit, but we only detect $n$-fold coincidence counts. So, $k$ photons may randomly be lost at the source, or at the detectors, or both can happen. Random loss at the single-photon source has been discussed in Ref \cite{Aaronson2016PRA}, and the output probability is given by $\Phi(A)=(1/\left|S\right|)\sum_{S}\left|\mathrm{Perm}(A_S)\right|^2$, where $A$ is a $n\times n$ submatrix, $S$ are all input combinations, and $|S|$ equals $\tbinom{n+k}{n}$. It is intuitive that, in this case, the output distribution is exactly the average of all possibilities that come from different input combinations. In this work, we give a clear formulation of the output probability when photons are lost at the detectors or both at single-photon sources and detectors \cite{Note1}. We give theoretical and numerical evidence that, path-independent loss, wherever it happens, is equivalent to a uniform loss at the single-photon source. Note that path-independent losses include all coupling loss through the optical path and the inhomogeneous loss at the detectors. So, for the following experiments, we just calculate $\Phi(A)$ and then modified it with the efficiency of the corresponding input ports and output ports, as in the theoretical distribution.

  \begin{figure*}
  \centering
  \includegraphics[width=0.85\textwidth]{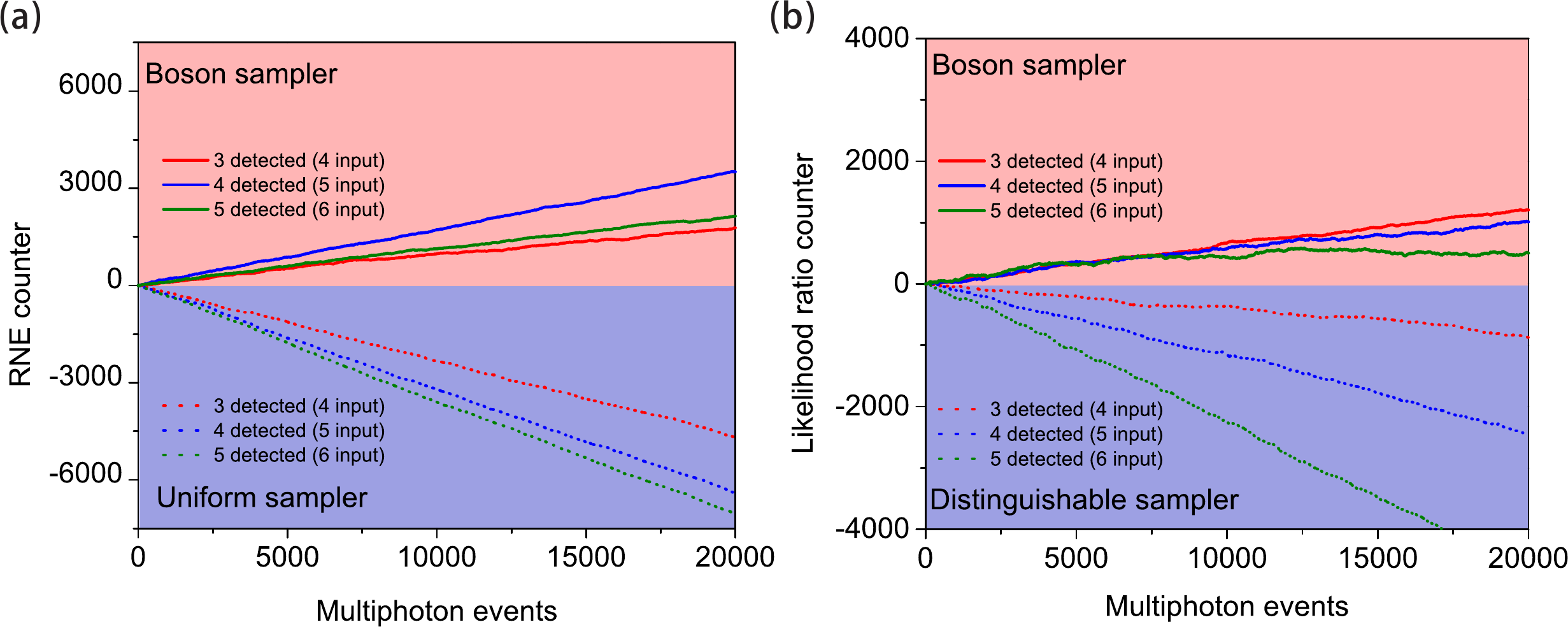}
  \caption{Validation of lossy boson sampling. (a) Application of extended RNE test to exclude the uniform distribution; (b) Extended likelihood ratio test to discriminate experimental data from a distinguishable sampler. The solid lines in (a) and (b) are tests applied on experimental data, and the dotted lines in (a) and (b) are tests applied on simulated data generated from a uniform sampler and a distinguishable sampler, respectively. The increasing difference between them indicate that experimental data are from a genuine boson sampler.}
  \label{Fig:3}
\end{figure*}

 \begin{figure*}
  \centering
  \includegraphics[width=1\textwidth]{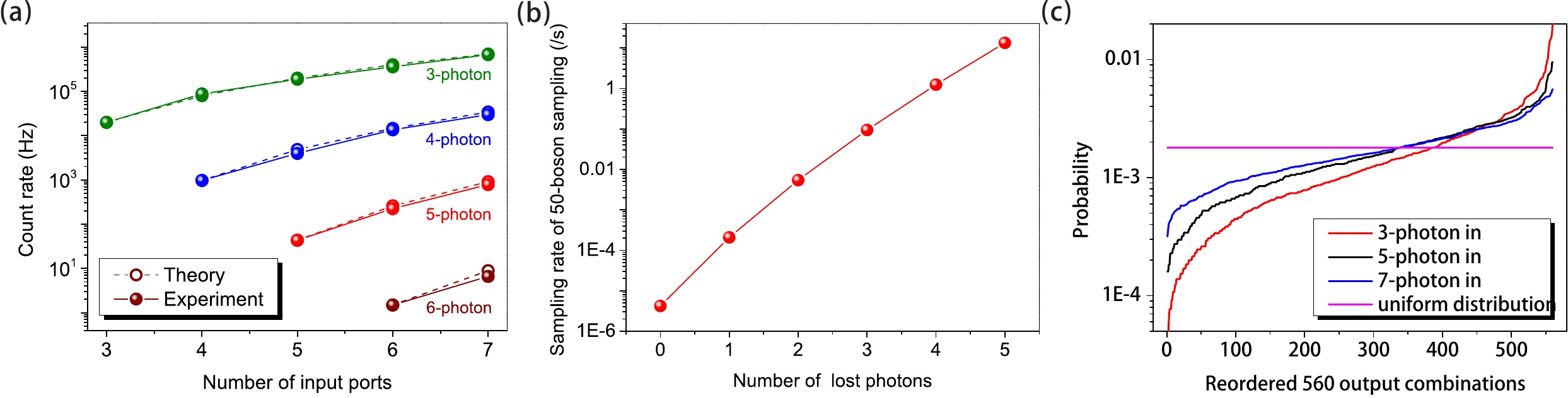}
  \caption{(a) The count rate comparison between experimental data and theoretical predictions. The excellent matches show that lossy boson sampling will have an exponential speed-up over the standard scenario by a factor of $\tbinom{n+k}{n}$. (b) The sampling rate of 50-photon boson sampling when the number of lost photons increase from zero to five. Note that, the assumed parameters of efficiency are 0.8 for single-photon sources, 0.9 for interferometer, and 0.9 for single-photon detectors, respectively. Allowing two photons lost will make this experiment feasible, while it's not realistic to perform standard boson sampling experiment with the same efficiency parameters. (c) Ascending ordered distributions of three-photon standard boson sampling (red), five-photon boson sampling with two photons lost (black) and seven-photon boson sampling with four photons lost (blue). If more photons are lost, the distribution will be closer to uniform distribution (pink).}
  \label{Fig:4}
\end{figure*}

We first studied boson sampling with one photon lost. We sent four, five and six photons into the sixteen-mode photonic network, but only extracted three-, four- and five-fold coincidence counts at the output ports of the interferometer. There are 560, 1820 and 4368 no-collision output combinations for four-, five- and six-boson sampling with one photon lost. The observed probability corresponding to each output combination is shown in the upper part of Fig.$\,$\ref{Fig:2}, while the bottom part is the theoretical probability given by calculating the corresponding submatrix.

A total of 402586 three-photon events, 198920 four-photon events, and 33587 five-photon events were obtained in an accumulation time of \SI{5}{s}, \SI{50}{s}, and \SI{150}{s}, respectively. To quantify the match between experimental distribution $(q_i)$ and theoretical distribution $(p_i)$, we calculated the total variation distance, defined as $D=(1/2)\sum_{i}\left|q_i-p_i\right|$, and the similarity, defined as $F=\sum_{i}\sqrt{q_ip_i}$. The obtained $D$ is 0.085(1), 0.106(2), 0.201(3), and the $F$ is 0.994(1), 0.989(2), 0.960(4) for four-, five-, and six-photon boson sampling with one photon lost, respectively. These two measures provide a first confirmation of the correct operation of the quantum devices.

 Next, we study the boson sampling with two photons lost. In our study, we sent five, six, and seven photons into the interferometer, and at the output, we registered three-, four-, and five-photon events (the detailed data of the distribution can be seen in \cite{Note1}). We found the measures of distance are 0.071(1), 0.097(2) and 0.178(3), and measures of similarity are 0.996(1), 0.992(2) and 0.967(4) for the five-, six-, and seven-photon boson sampling with two photons lost, respectively.

 To give further supporting evidence that our experimental results are from genuine boson sampling, we applied several statistical tests to rule out possible alternative hypotheses. We first excluded the hypothesis that distinguishable single photons or spatial-mode mismatched interferometers were used, by performing a new version of a standard likelihood ratio test \cite{Spagnolo2014NP,Cover2006}. The correctness of this method was shown by the simulated results \cite{Note1}. Figure$\,$\ref{Fig:3}(a) shows an increasing difference between experimental data and simulated data by distinguishable bosons. The Aaronson and Arkhipov test is designed to distinguish boson sampling from a uniform distribution \cite{Aaronson2014uniform,Spagnolo2014NP}. Here, we extend it to the lossy boson sampling \cite{Note1}, and Fig.$\,$\ref{Fig:3}(b) clearly shows the difference between boson sampler and uniform sampler.

 Like scattershot boson sampling \cite{Lund2014PRL, Bentivegna2015SA}, lossy boson sampling is expected to show an speed-up over standard boson sampling. In our experiments,the rate of three-photon boson sampling is \SI{19.9}{KHz}, which is over $10^5$ times faster than all previous boson sampling experiments based on the SPDC sources. When we increase the number of lost photons from one to four, the rates change to \SI{87.8}{KHz}, \SI{187.9}{KHz}, \SI{357.7}{KHz}, \SI{673.0}{KHz}, which are 4.4, 9.4, 17.9, 33.8 times faster than standard boson sampling, respectively. In Fig.$\,$\ref{Fig:4}(a), we present the sampling rates of all conditions up to six-photon boson sampling, and the data is in excellent agreement with theoretical prediction. Next, we will discuss 50-photon boson sampling with photon loss. The assumed realistic parameters of efficiency are 0.8 for single-photon sources, 0.9 for a interferometer by our integrated bulk optics approach (including the efficiency of coupling the photons into single-mode fibers), and 0.9 for single-photon detectors. In this condition, the rate for standard 50-photon boson sampling is $10^{-6}$ Hz. However, if we have 52 input ports, and allow two photon loss, the rate becomes $\sim$\SI{0.005}{Hz}, which is already feasible to do such an experiment. In Fig.$\,$\ref{Fig:4}(b), we present sampling rate of lossy boson sampling with up to five photons lost, which is likely to be a scalable approach to demonstrate quantum supremacy.

 Last but not the least, we discuss the open theoretical questions in lossy boson sampling. As discussed in Ref. \cite{Aaronson2016PRA}, if $k$ is fixed, lossy boson sampling remains in the same complexity class as standard boson sampling in the limit n$\rightarrow$$\infty$. However, it's still an open question what complexity it retains in a realistic loss regime, say, $k=\sqrt{n}$ or $k=\ln{(n)}$. What will happen when the number of lost photons $k$ increases? Intuitively, it will make the problem easier since the dimension of the submatrices become smaller. On the other hand, since the sum is taken over $\tbinom{n+k}{n}$ input combinations, the distribution will be much flatter than standard boson sampling, so there may exist some approximate algorithms to simulate this distribution. As an example, we sent up to seven photons into interferometer and only detected three-fold coincidence counts. Fig.$\,$\ref{Fig:4}(c) is the reordered distribution (ascending order) of three-photon boson sampling with three (red), five (black) and seven (blue) input ports, respectively. It shows that, when the number of lost photons increases, the distribution will be closer to the uniform distribution (pink). Where is the threshold value of $k$ so that lossy boson sampling is intractable for classical computers is an important open question, since $k$ is the key for the least demanding efficiency. Our experiment shows a possible way to increase the multiphoton boson sampling efficiency which could be helpful to achieve quantum supremacy with single photons and linear optics. We hope that our experiment will inspire more work on lossy boson sampling.

\
\begin{acknowledgments}
We thank S. Aaronson, B. Sanders and J.-Z. Wu for helpful discussions. This work was supported by the National Natural Science Foundation of China, the Chinese Academy of Science, the Science and Technology Commission of Shanghai Municipality, the National Fundamental Research Program, the State of Bavaria, and the US National Science Foundation.
\end{acknowledgments}

\rule[-10pt]{8.3cm}{0.1em}

\appendix
\section{Supplementary Material}

\subsection{Experimental details on single-photon source}
We directly refer to Ref \cite{Wang2017NP} and its supplementary information for all experimental details on the quantum dot single-photon source.

\subsection{Details on demultiplexers}
Under pulse excitation, a quantum dot emits a long stream of single photons. In order to separate these single photons into several different spatial modes, fast optical switches are employed in our experiments. We note that our method eliminates the need for growth and control of homogeneous quantum dots, which was a known challenge due to the self-assembly process.

\begin{figure*}
  \centering
  \includegraphics[width=0.95\textwidth]{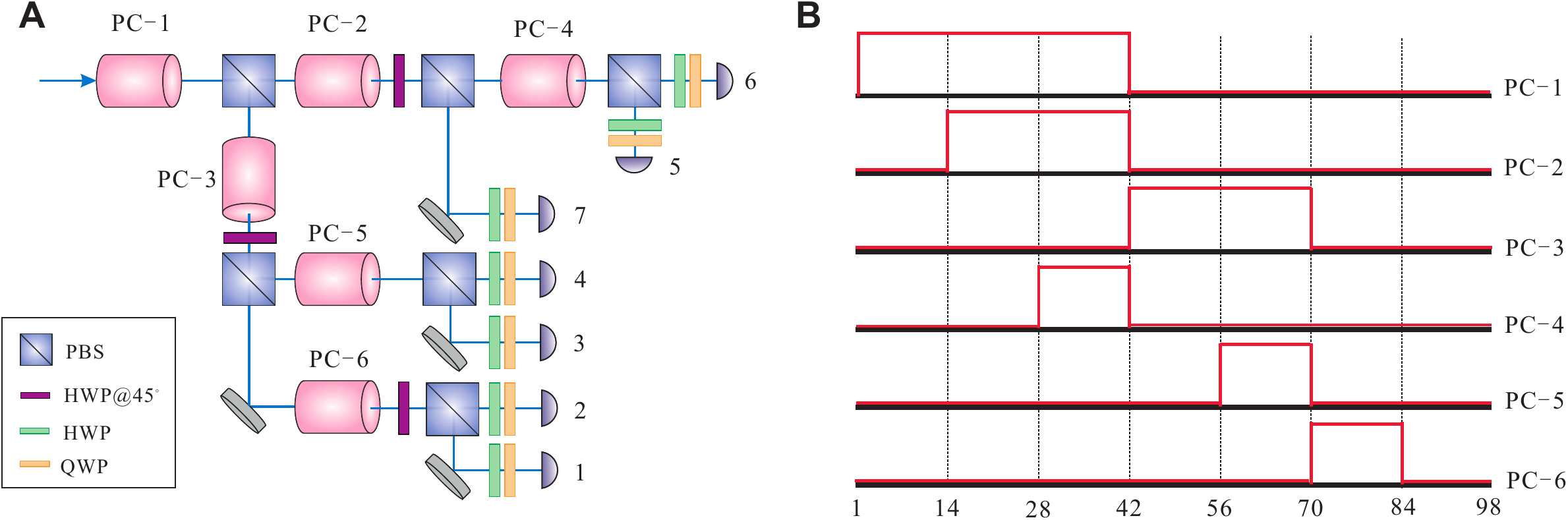}
  \caption{Details of demultiplexers (A) and the electric signals (B).}
  \label{fig:S1}

\end{figure*}

Under a half-wave voltage ($\sim$\SI{1800}{V}), the Pockels cell (PC) rotates the photon's polarization from $V$ (or $H$) to $H$ (or $V$), but doesn't change it when no electric signal applied (We refer to Ref \cite{Wang2017NP} for more details for the PC). All PCs operated at a repetition rate of \SI{0.775}{MHz}, 1/98 of the \SI{75.95}{MHz} repetition rate of our pulsed laser. Hence, for every cycle, we intend to separate 98 pulses into 7 spatial modes. Fig. \ref{fig:S1}A is the configuration of our demultiplexers, and Fig. \ref{fig:S1}B shows the electric pulses applied to drive the PCs. As an example, we just explain how we get 14 pulses at the port 7. First, single photons are all prepared with $V$ polarization. Driven by the signal shown in Fig. \ref{fig:S1}B, PC-1 rotates the $1_{st}$-$42_{nd}$ pulses from $V$ to $H$, and then PBS behind the PC reflects the $V$ polarized photons while transmits the H polarized photons. Next, PC-2 changes the polarization of $15_{th}$-$42_{nd}$ pulses from $H$ to $V$. For some special reasons, a $45^{\circ}$ half wave plate is used to reverse the polarization, so $1_{st}$$-14_{th}$ pulses are directly reflected by a PBS to the port 7. With a similar process, we eventually obtain 7 spatially-separate single-photon pulses.

\begin{figure}
  \centering
  \includegraphics[width=1\columnwidth]{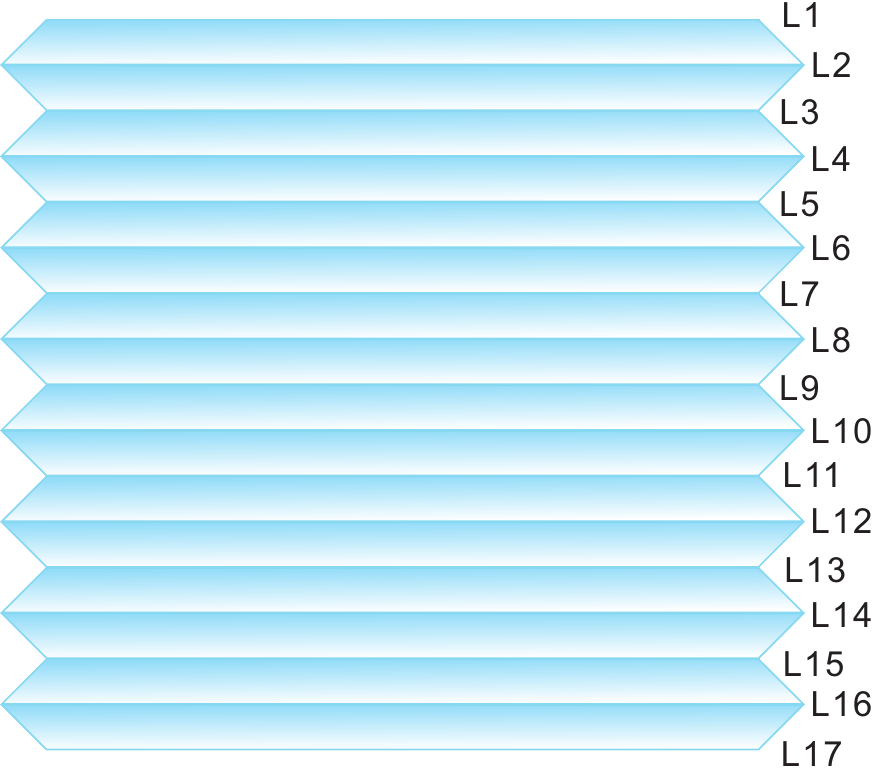}
  \caption{~$16\times16$ mode ultra-low-loss photonic circuits.}
  \label{fig:S2}

\end{figure}

\begin{figure}
  \centering
  \includegraphics[width=1\columnwidth]{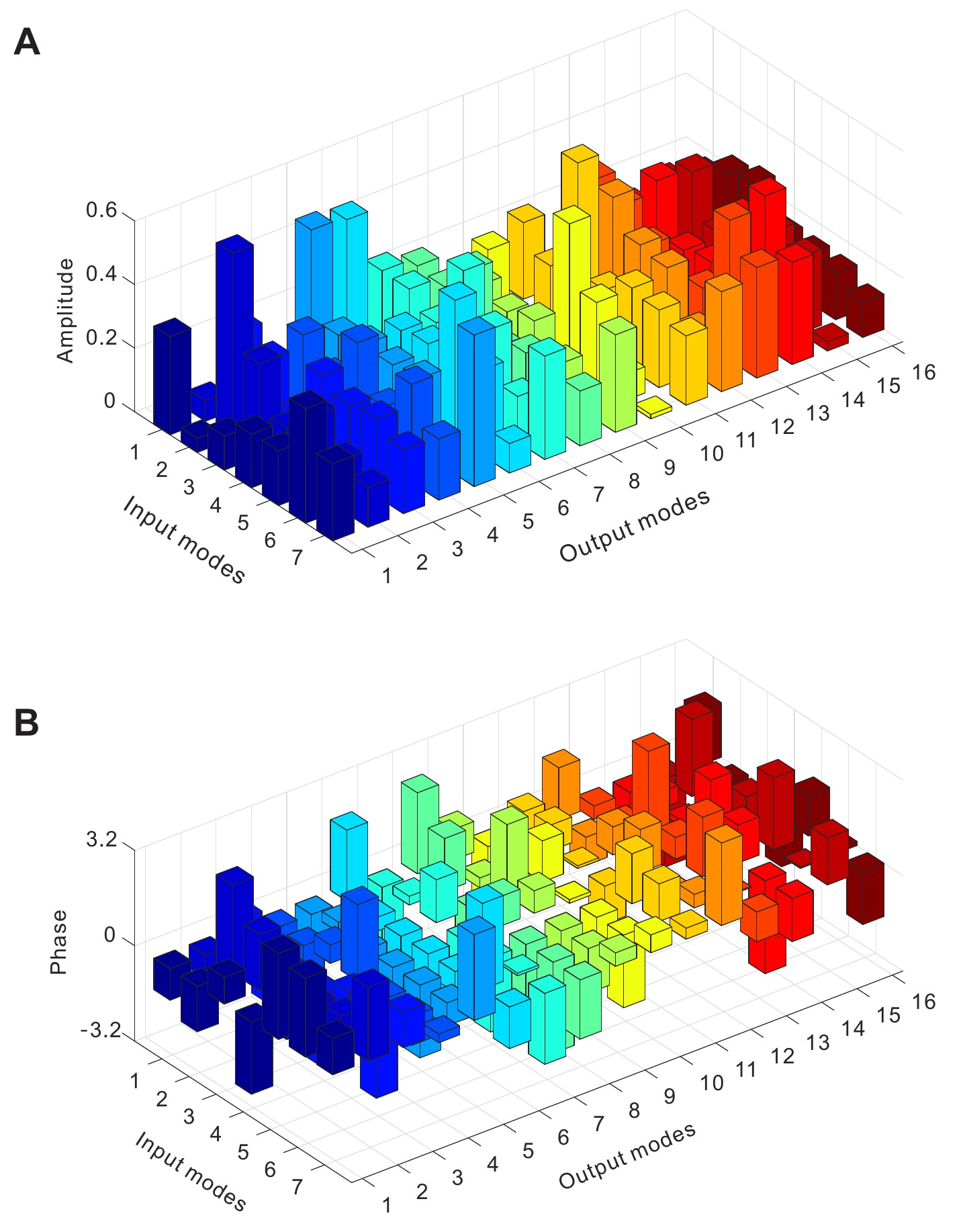}

  \caption{Measured elements of unitary matrix.}
  \label{fig:S3}

\end{figure}
To ensure that these photons from different paths can overlap on the interferometer, we use single-mode fibers of various length of \SI{2.6}{m}, \SI{39}{m}, \SI{76}{m}, \SI{113}{m}, \SI{150}{m}, \SI{186}{m}, \SI{223}{m} to compensate their relative time delay. In addition, before feeding them into the interferometer, translation stages with a traveling range of \SI{25}{mm} are employed to finely adjust the temporal delay. The measured efficiency of each channel is $\sim$85$\%$ on average. The main losses come from coupling and propagation loss in the long fibers. Note that, losses at the demultiplexers are easy to take into account in lossy boson sampling.

In order to get correct coincidence counts, we have to synchronize the single-photon pulses with the PCs driving signals and the 64-channel coincidence counting unit (CCU). To this end, the electric signal converted by a small fraction of the pulsed laser is fed into a field programmable gate array (FPGA) board which serves as a frequency divider with a division factor of 98 to produce two-channel trigger signals. One of them is used as a trigger of the PC's, and the other is fed into the CCU as an external clock source. Our home-built 64-channel CCU only registers the multiphoton events we want, and we can see in real-time the results of our experiments.

\subsection{Fabrication and characterization of the photonic network}

The Photonic network, which serves as a unitary transformation to input single-photon Fock states, is the key ingredient in boson sampling experiments. In this work, we design and fabricate a new photonic network, which combines a near-unity transmission rate, phase stability, Harr randomness, robustness to the environmental fluctuations, and high level of integration. As shown in Fig. \ref{fig:S2}, the square-shaped \cite{Clements2016} 16$\times$16 mode circuit consists of 113 beam splitters and 14 mirrors. Such a circuit is made by 16 trapezoid-shaped fused quartz plates with a size of 50.91mm$\times$45.25mm$\times$4.00mm. Firstly, each trapezoid is cut and finely polished with a dimensional tolerance below ~\SI{5}{\micro\metre}, and a angle tolerance below \SI{24}{\micro\radian}. Then, L1 and L17 are total-reflecting coated, and L2-L16 are optically coated with polarization-dependent beam-splitting thin films. The transmission to reflection ratios are 0.65:0.35 and 0.85:0.15 for $H$ and $V$ polarization, respectively. Next, all 32 outer surfaces (16 for input ports and 16 for output ports) are antireflection coated to further improve the transmission efficiency. Finally, the 16 trapezoids are bond together one-by-one via intermolecular Van der Waals forces.

The unitary transformation, implemented by our 16$\times$16 mode interferometer, is experimentally measured by the method described in Ref \cite{Keshari2016}. The measured elements are shown in Fig. \ref{fig:S3}. To characterize the spatial-mode overlap in our 16-mode interferometer, we applied a coherent laser to do Mach-Zehnder-type coherence measurements. We chose any two ports to do such a measurement (after finely tune the output intensity to be equal); the overlap always exceeded 99.9$\%$ with single-mode fiber coupling. Next, we fed a laser with an intensity of $I_{total}$ into one of the input ports, and measured the intensity at all output ports, denoted as
$I_1$, $I_2$, $\cdots$, $I_16$, respectively. Thanks to the antireflective coating, the efficiency
$\eta=\sum_{i=1}^{16}I_i/I_{total}$ is always determined to be above 99$\%$. Also, the phase of this square-shaped interferometer can remain stable for weeks, like our triangle-shaped one (used in Ref \cite{Wang2017NP}).

\subsection{Analysis of the lossy boson sampling}
\subsubsection{Losses at the single-photon source}

Here, losses at the single-photon source include all transport losses and coupling losses before feeding into the 16$\times$16 mode interferometer, and we assume all losses happen before the interferometer. For simplicity, first, we suppose the uniform losses happen at all input ports. The losses can be represented by an operator which can be written as $L_{in}=\sqrt{\xi}I$, where $\xi$ is the uniform efficiency of each channel, and $I$ is an identity. We denote the transformation matrix of our interferometer as $U$. Hence the final state is $\left|\psi\right\rangle_{out}=UL_{in}\left|\psi\right\rangle_{in}=\sqrt{\xi} U\left|\psi\right\rangle_{in}$. In this case, Aaronson and Brod \cite{Aaronson2016PRA} have shown that the probability of outcome is $\Phi(A)=(1/\left|S\right|)\sum_{S}\left|Perm(A_S)\right|^2$ (see main text), here the ~$A$ is a submatrix of ~$\sqrt{\xi} U$. In our experiments, we only detect non-collision multiphoton events, the $\xi$ was eliminated during probability normalization. In this case, we can ignore $\xi$ totally.

\begin{figure*}
  \centering
  \includegraphics[width=0.95\textwidth]{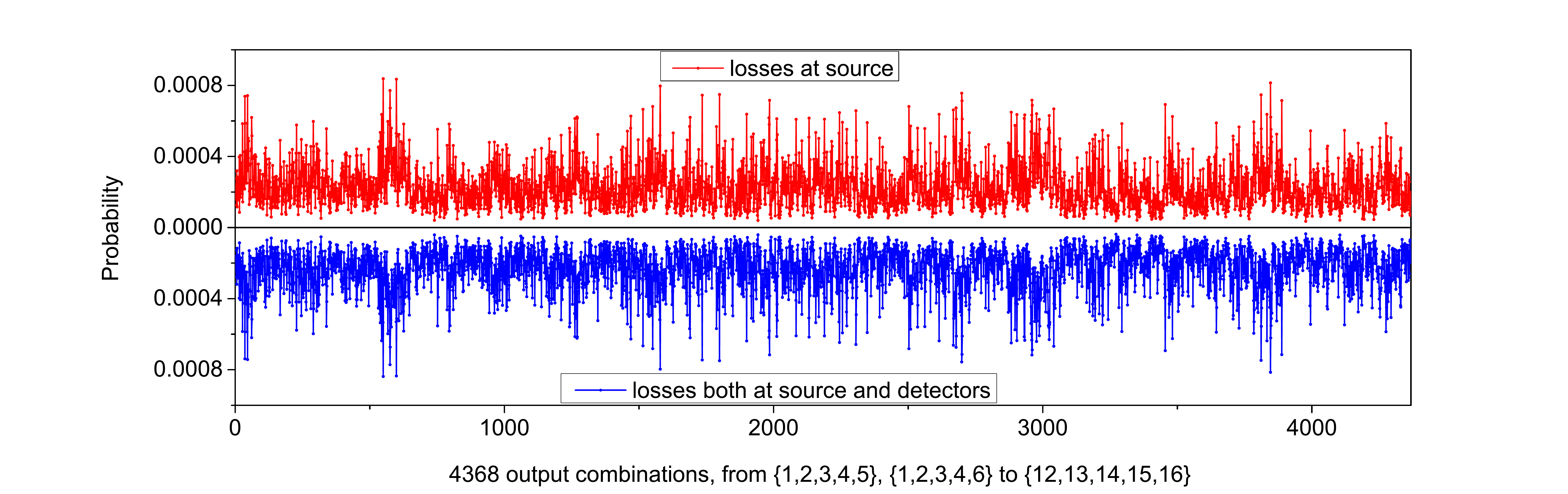}
  \caption{Comparison of the two cases---losses at the single-photon sources and losses at both single-photon source and detectors. We assume that seven photons was fed into the system, but we detect all five-fold coincidence counts. If all losses are uniform for different pathes, we can see the simulated distributions are identical.}
  \label{fig:S4}

\end{figure*}
\begin{figure}
  \centering
  \includegraphics[width=1\columnwidth]{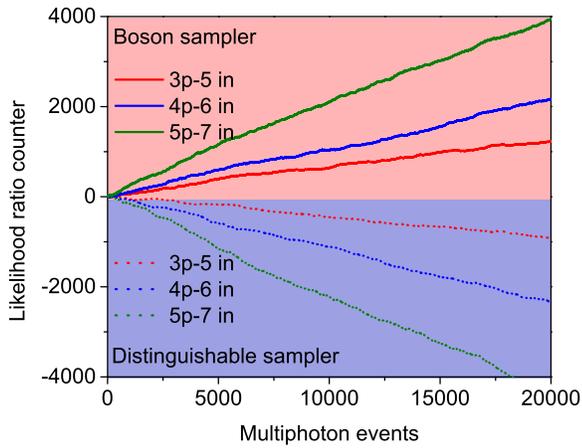}
  \caption{Simulated extended likelihood ratio test of boson sampling with one photon lost.}
  \label{fig:S5}

\end{figure}

\begin{figure}
  \centering
  \includegraphics[width=1\columnwidth]{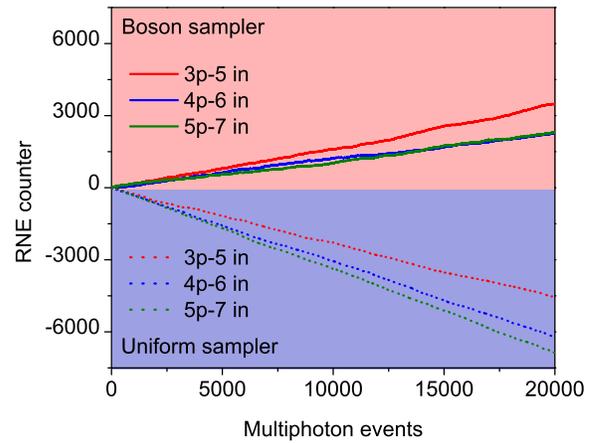}
  \caption{Simulated extended RNE test of boson sampling with one photon lost.}
  \label{fig:S6}

\end{figure}

However, in our experiments we demultiplex single photons into seven input ports.
Due to the different length of the fibers and different coupling efficiency,
we can't ensure exactly the same efficiency for all seven channels.
The operator can be written as
$L_{in}^{\prime}=diag\left\{\sqrt{\xi_1}, \sqrt{\xi_2}, \ldots, \sqrt{\xi_n}\right\}$,
where $\xi_i$ is the efficiency of the ~$i-th$ input port, and ~$n$ is the number of input ports.
Therefore, the final state is
$\left|\psi^{\prime}\right\rangle_{out}=UL_{in}^{\prime}\left|\psi\right\rangle_{in}$ and $\Phi(A)=(1/\left|S\right|)\sum_{S}\left|Perm(A_S)\right|^2=(1/\left|S\right|)\sum_{S}(\prod_{\sigma\in S}\xi_{\sigma})\left|Perm(A_S)\right|^2$,
since $Perm(UL_{in})=\prod_{i=1}^n\sqrt{\xi_i}Perm(U)$.
Therefore, in this case, we only need to modify the probability
$\left|Perm(A_S)\right|^2$
by the factor
$\prod_{\sigma\in S}\xi_{\sigma}$
for every input combination $S$.

\subsubsection{Losses at the detectors}
As defined in the main text, losses at the detectors consist of all losses after the interferometer, such as transport loss, coupling loss, and detection loss. In this case, losses can be represented as a diagonal operator $L_{out}$. If losses are uniform for all output ports, it¡¯s easy to see that it is equivalent to the uniform losses at the source. Actually, the efficiency of 16 output ports are totally different mainly due to the difference among detectors¡¯ efficiency. Similarly, the operator for losses can be written as $L_{out}^{\prime}=diag\left\{\sqrt{\varepsilon_1}, \sqrt{\varepsilon_2}, \ldots, \sqrt{\varepsilon_m}\right\}$, where the $\varepsilon_i$ is the efficiency of $i-th$ output port, and $m$ is the number of output ports. In this case, the output state is $\left|\psi^{\prime}\right\rangle_{out}=L_{out}^{\prime}U\left|\psi\right\rangle_{in}$. Then, the output probability is $\Phi(A^{\prime})=(1/\left|S^{\prime}\right|)\sum_{S^{\prime}}\left|Perm(A^{\prime}_{S^{\prime}})\right|^2$ if photons are lost only at the detectors. In this case, $A^{\prime}$ is a n$\times$n submatrix, since all $n$ photons transport the interferometer, but we only register all $(n-k)$-fold coincidence counts.And $S^\prime$ are all output combinations, which include all collision modes, and $\left|S^\prime\right|$ is equal to $\tbinom{m+k-1}{k}$. Note again that the sum is taken over output combinations. For example, we fed 5 photons into the 16$\times$16 mode interferometer, but only detected 4 photons at the output combination $\left\{1,2,3,4\right\}$. One photon may lost at the ports 1, 2, $\ldots$, 16, so the output combinations $S^\prime$ are $\left\{1,1,2,3,4\right\}$, $\left\{1,2,2,3,4\right\}$, $\ldots$, $\left\{1,2,3,4,16\right\}$, and $\left|S^{\prime}\right|$ equals to 16. Due to the fact that $L_{out}^{\prime}$ is a diagonal matrix, we can get $Perm(L_{out}^{\prime}U)=\prod_{i=1}^m\sqrt{\varepsilon_i}Perm(U)$. so if losses happen only at the detectors, the probability is $\Phi(A^{\prime})=(1/\left|S^{\prime}\right|)\sum_{S^{\prime}}\left|Perm(A^{\prime}_{S^{\prime}})\right|^2=(1/\left|S^{\prime}\right|)\sum_{S^{\prime}}\prod_{\sigma\in S^{\prime}}\varepsilon_{\sigma}\left|Perm(U)\right|^2$. Hence, if all losses happen at the detectors, we just need to multiply a factor $(\prod_{\sigma\in S^{\prime}}\varepsilon_{\sigma})$ for every output combination $S^\prime$.

\subsubsection{Losses both at single-photon sources and detectors}
Thanks to the ultra-low-loss photonic circuits we used in our experiments, we don't need to take the circuits' loss into account since it has a transmission rate above 99$\%$. But the unavoidable losses at the single-photon sources and detectors both happen in reality, even we don't know where photons were lost at all. In this case, we combine the results given in the first two parts in this section. Here, we give the probability as $\Phi(A^{\prime\prime})=\sum_S\sum_{S^{\prime}}\left|Perm(A^{\prime\prime}_{S^{\prime}})\right|^2$ for the uniform losses case, where S are the input combinations when $k_1$ photons are lost at the single-photon source, while $S^\prime$ are the output combinations when $k_2$ photons are lost at the detectors, and $k_1+k_2=k$. As an example, we assume that seven photons were fed into interferometer with an input combination $\left\{1,2,3,4,5,6,7\right\}$. If one photon lost at the source $(k_1=1)$, and one photon lost at detectors $(k_2=1)$, we only could detect five-fold coincidence counts. Hence, $S$ consists of $\left\{1,2,3,4,5,6\right\}$, $\left\{1,2,3,4,5,7\right\}$, $\ldots$, $\left\{2,3,4,5,6,7\right\}$; $A_{S^{\prime}}$ is a $6\times6$ submatrix. If we detected a 5-fold even at the output combination $\left\{1,2,3,4,5\right\}$, then the probability should take sum over $\left\{1,1,2,3,4,5\right\}$, $\left\{1,2,2,3,4,5\right\}$, $\ldots$, ~$\left\{1,2,3,4,5,16\right\}$ (these are the elements of $S^\prime$). We note that, in this case, the $\Phi(A^{\prime\prime})$ is equivalent to the $\Phi(A)$
defined in the first part. We give numerical evidence in Fig. \ref{fig:S4}, and A is distribution given by $\Phi(A)$, while B is by $\Phi(A^{\prime\prime})$, they are totally identical. For the nonuniform loss case, we just need to modify the probability with a factor caused by different efficiencies at the input and output ports (see the first two parts in this section).

\subsection{Validation}
\subsubsection{Extended likelihood ratio test}
The standard likelihood ratio test was originally designed to distinguish boson sampling from distinguishable photon sampling. Here, we extend it to the lossy boson sampling. Let $P_k^{indis}$ and $P_k^{dis}$ denote the probabilities associated with indistinguishable and distinguishable photons for the observed event $k$, respectively. We first define two probabilities as $P_k^{indis}=\sum_S\left|Perm(A_S)\right|^2$, and $P_k^{dis}=\sum_SPerm(\left|A_S\right|^2)$, and then calculate the estimator $L_k=P_k^{indis}/P_k^{dis}$. The counter $C$ is initialized to 0 and then updates as follows:

\begin{equation}
  C: = \begin{cases}
  C, & a_1<L_k<1/a_1\\
  C+1 & 1/a_1\leq L_k<a_2\\
  C+2 & L_k\geq a_2\\
  C-1 & 1/a_2\leq L_k<a_1\\
  C-2 & L_k\leq 1/a_2\\
  \end{cases} .
\end{equation}

Finally, after $N$ events, if $D\geq0$, the test decides that the outcomes are from boson sampling data, otherwise the outcomes are from distinguishable photon sampling.

To show the feasibility of the method, we simulated the update of the counter C, Fig. \ref{fig:S5} shows the simulated results; it showed the clear difference between boson sampling and distinguishable photon sampling. In our test, we set $a_1=0.9$ and $a_2=1.5$. Our validation results using experimental data was shown in the main text.

\begin{figure*}
  \centering
  \includegraphics[width=0.95\textwidth]{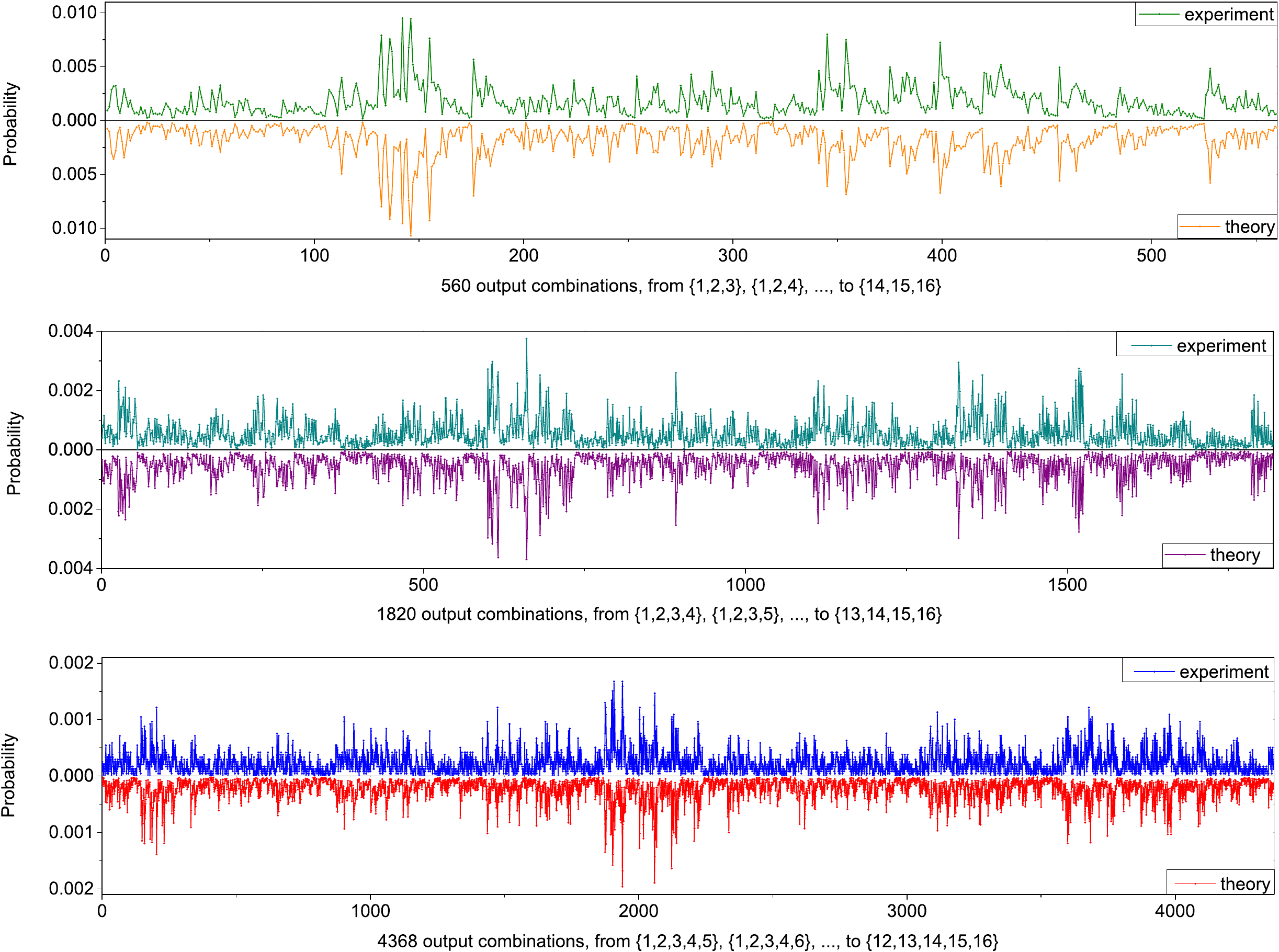}
  \caption{Distribution of boson sampling with two photons lost.}
  \label{fig:S7}

\end{figure*}

\begin{figure*}
  \centering
  \includegraphics[width=0.85\textwidth]{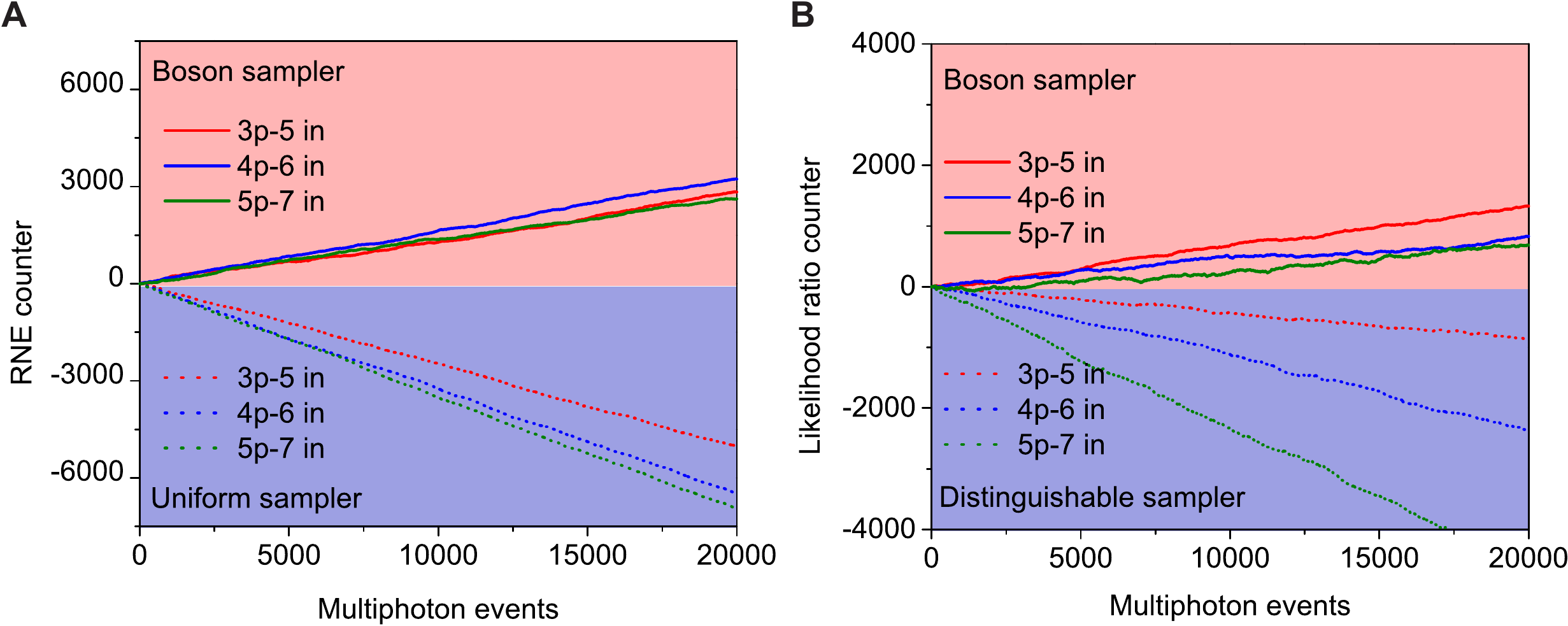}
  \caption{Experimental validation of boson sampling with two photons lost.}
  \label{fig:S8}

\end{figure*}

\subsubsection{Extended Aaronson and Arkhipov test (or extended RNE test)}

RNE method, proposed by Aaronson and Arkhipov, is used to test data against the uniform sampling. Again, we extend it to the lossy scenario. The Extend RNE test works as following: Firstly, the counter $C$ is initialized to 0. For each measured event, we calculate the estimator $P_k=\sum_S(\prod_i\sum_j\left|A_{i,j}\right|_S^2)$ for the corresponding submatrix $A$ of $U$. Then, the counter $C$ updates according to the following rules:

\begin{equation}
  C:=\begin{cases}
  C+1, & \text{ if} \,P_k>(n/m)^n\\
  C-1 & \text{ else }\\
  \end{cases}.
\end{equation}

For boson (uniform) sampling, counter $C$ increases (decreases) almost monotonically, which allows us to distinguish these two hypotheses. Also, our simulation results (Fig. \ref{fig:S6}) give numerical evidence that this method works, and we then applied it to our experimental data (see main text).

\subsection{More experimental results}

Here, we show the experimental results of boson sampling with two photons lost (Fig. \ref{fig:S7}). From the top to bottom are five-, six- and seven-photon boson sampling with two photons lost, respectively. The upper part is the experimental result and the bottom part is the theoretical distribution. We found the measures of distance between them are 0.071, 0.097 and 0.178, and the measures of fidelity are 0.996, 0.992 and 0.967, respectively. Fig.\ref{fig:S8} is the validation of the experimental results. We applied the extended RNE test (Fig. \ref{fig:S8}A) to rule out the uniform distribution, and the extended likelihood ratio test (Fig. \ref{fig:S8}B) to exclude the distinguishable distribution. Our tests showed that our experimental data are from a genuine boson sampler.

\bibliography{lost}

\end{document}